\documentclass[%
 reprint,
 superscriptaddress,
 amsmath,amssymb,
 aps,
]{revtex4-2}

\usepackage{graphicx}
\usepackage{dcolumn}
\usepackage{bm}
\usepackage{siunitx}
\usepackage{hyperref}
\usepackage[mathlines]{lineno}
\usepackage{color}

\begin{document}

\preprint{APS/123-QED}

\title{Integrating Classical and Quantum Software for Enhanced Simulation of Realistic Chemical Systems}

\author{Tomoya Shiota}
\email{shiota.tomoya.ss@gmail.com}
\affiliation{Graduate School of Engineering Science, The University of Osaka, 1-3 Machikaneyama, Toyonaka, Osaka 560-8531, Japan}
\affiliation{Center for Quantum Information and Quantum Biology, The University of Osaka, 1-2 Machikaneyama, Toyonaka 560-8531, Japan} 

\author{Klaas Gunst}
\affiliation{Quantum Simulation Technologies, Inc., Cambridge, MA 02139}

\author{Toshio Mori}
\affiliation{Center for Quantum Information and Quantum Biology, The University of Osaka, 1-2 Machikaneyama, Toyonaka 560-8531, Japan}

\author{Toru Shiozaki}
\affiliation{Quantum Simulation Technologies, Inc., Cambridge, MA 02139}

\author{Wataru Mizukami}
\email{mizukami.wataru.qiqb@osaka-u.ac.jp}
\affiliation{Graduate School of Engineering Science, The University of Osaka, 1-3 Machikaneyama, Toyonaka, Osaka 560-8531, Japan}
\affiliation{Center for Quantum Information and Quantum Biology, The University of Osaka, 1-2 Machikaneyama, Toyonaka 560-8531, Japan} 

\date{\today}

\begin{abstract}
We demonstrate the feasibility of quantum computing for large-scale, realistic chemical systems through the development of a new interface using a quantum circuit simulator and CP2K, a highly efficient first-principles calculation software. Quantum chemistry calculations using quantum computers require Hamiltonians prepared on classical computers. Moreover, to compute forces beyond just single-point energy calculations, one- and two-electron integral derivatives and response equations are also to be computed on classical computers.
Our developed interface allows for efficient evaluation of forces with the quantum-classical hybrid framework for large chemical systems. We performed geometry optimizations and first-principles molecular dynamics calculations on typical condensed-phase systems. These included liquid water, molecular adsorption on solid surfaces, and biological enzymes. In water benchmarks with periodic boundary conditions, we confirmed that the cost of preparing second-quantized Hamiltonians and evaluating forces scales almost linearly with the simulation box size. This research marks a step towards the practical application of quantum-classical hybrid calculations, expanding the scope of quantum computing to realistic and complex chemical phenomena.
\end{abstract}

\maketitle

\section{\label{sec:level1}Introduction}

Computational chemistry has emerged as an indispensable tool for understanding chemical phenomena that occur in various phases~\cite{jensen2017introduction, cao2019quantum}. Today, quantum chemical calculations, particularly those based on first-principles, facilitate (semi-)quantutitative simulations of nanoscale world. Recently, quantum computers have garnered significant attention for their potential to advance quantum chemical calculations~\cite{romero2018strategies,cao2019quantum, McArdle_2020,Cerezo_2021, Bauer_2020, Motta_2021, fedorov2022vqe}. Over the past decade, especially in the last five years, a diverse array of quantum algorithms for quantum chemistry has been developed~\cite{Peruzzo_2014, Kandala_2017, Babbush_2018, Babbush_2018_2, Berry_2019, Low_2019, Mizukami_2020, von2021quantum, Motta_2021_low, Lee_2021, Omiya_2022, kohda2022quantum, arrazola2022universal,PhysRevResearch.4.043210, Sugisaki_2022, arrazola2022universal, erhart2022constructing, malone2022towards, hohenstein2023efficient, Nakagawa_2023, loipersberger2023accurate, Mitarai_2023, rubin2023faulttolerant, Ivanov2023periodic, kanno2023quantum, nakagawa2023adapt, yoshida2024solvent, yoshida2024ab, oumarou2024accelerating, ohgoe2024demonstrating, erhart2024coupled, erhart2024chebyshev, cortes2024fault}, with numerous demonstrations using actual quantum devices~\cite{Peruzzo_2014, Kandala_2017, google2020hartree, ohgoe2024demonstrating, robledo2024chemistry}.

Despite these advancements, the scope of quantum chemical calculations employing quantum algorithms has largely been confined to small molecules in the gas phase or crystals with small unit cells, barring a few exceptions. While calculations on surface chemical reactions~\cite{gujarati2023quantum}, thermally activated delayed fluorescence (TADF) systems ~\cite{gao2021applications}, and per- and poly-fluoroalkyl substances (PFAS)~\cite{dimitrov2023pushing} have been reported, these still represent relatively small molecular systems by today’s quantum chemistry standards~\cite{KRESSE199615, guidon2008ab, frisch2016gaussian, giannozzi2017advanced, sun2018pyscf, shiozaki2018bagel, neese2020orca, smith2020psi4, kuhne2020cp2k, barca2020recent, shiota2020microscopic, seritan2021terachem, li2024introducing, li2024accurate, bussy2024efficient}. Moreover, the majority of these studies have been limited to single-point energy calculations using geometries structures pre-optimized with density functional theory (DFT).

Notably, there are few studies applying quantum algorithms to practical materials for common quantum chemical tasks such as transition state searches, geometry optimizations, and ab initio molecular dynamics (AIMD) simulations~\cite{fedorov2021ab, gujarati2023quantum, hohenstein2023efficient}. The work of Hohenstein et al. stands as a rare exception, yet even their results pertain to finite systems~\cite{hohenstein2023efficient}. To date, no reports exist of AIMD simulations or geometry optimizations using quantum algorithms for condensed-phase systems under periodic boundary conditions (PBC).

This current state of affairs stems not from limitations in quantum algorithms, but rather from constraints in classical computing. Electronic structure calculations using quantum computers, whether employing quantum phase estimation (QPE)~\cite{Aspuru_Guzik_2005, Babbush_2018} or the variational quantum eigensolver (VQE)~\cite{Peruzzo_2014, grimsley2019adaptive, Cerezo_2021, fujii2022deep}, are inherently quantum-classical hybrid computations. These methods virtually utilize second-quantized Hamiltonians prepared on classical computers. Even with a limited number of qubits, quantum algorithms can be applied to larger systems through active space approximations, provided that an appropriate active orbital space Hamiltonian is available.
The computational cost of generating electronic Hamiltonians formally scales as $O(N^5)$, or $O(N^4)$ when employing the active space approximation. Furthermore, force calculations essential for geometry optimization and AIMD simulations necessitate the computation of nuclear coordinate derivatives of one- and two-electron integrals and orbital response terms~\cite{osamura1982generalization}. The cost of these calculations for large-scale systems on classical computers is substantial, particularly under PBC~\cite{ferrero2008coupled, paier2009accurate, erba2022crystal23}.

Addressing the classical computational cost issue in quantum chemical calculations using quantum algorithms is crucial for applying quantum computers to systems of interest in materials science. Tackling this challenge requires the use of software implementing more sophisticated algorithms on the classical side.

In this study, we have developed an interface between the large-scale quantum chemistry program package CP2K version 8.2~\cite{kuhne2020cp2k} and quantum computation. While Battaglia et al.~\cite{battaglia2024general} have already developed a interface between CP2K and a quantum computing software Qiskit~\cite{qiskit2024}, their research primarily focused on realizing single-point energy calculations, particularly in the framework of a wavefunction-in-DFT embedding method~\cite{lee2019projection}. Similarly, Lukas Schreder and Sandra Luber have built an interface between CP2K and OpenMolcas~\cite{fdez2019openmolcas}, implementing the wavefunction-in-DFT embedding method.~\cite{schreder2024implementation} In contrast, our interface emphasizes force calculations in quantum chemical computations using quantum algorithms. Consequently, we have achieved, for the first time using quantum algorithms in first-principles calculations, under PBC, the exploration of molecular adsorption structures on solid surfaces and AIMD simulations of water in the liquid phase.

The remainder of this paper is organized as follows. In Section II, we describe the interface for the quantum-classical hybrid algorithm for quantum chemistry integrating CP2K and quantum circuit simulators. In Section III, we demonstrate the application of our developed interface in case of representative systems in gas, solid, and liquid phases. Further, we discuss the computational efficiency in terms of the system size. Section IV concludes with a summary of our study and its implications for the advancement of quantum computing in computational chemistry.

\begin{figure*}[]
\includegraphics{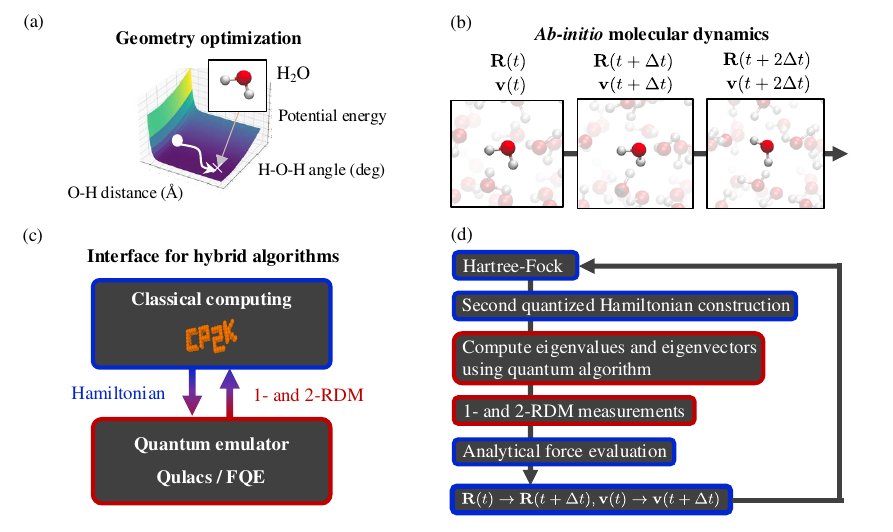}
\caption{\label{fig:1} (a) Geometry optimization for a water molecule, illustrating the potential energy surfaces as functions of the O-H bond distance and H-O-H bond angle. Geometry optimization follows the forces from the initial geometry, indicated by a white circle, to the equilibrium geometry, marked via a cross. (b) \textit{Ab initio} molecular dynamics (MD) simulations, highlighting the finite-temperature dynamical trajectories of liquid water. In MD, the evaluation of velocity $\mathbf{v}$, which is derived from the forces based on the gradient of the total energy $E$ with respect to atomic coordinates $\mathbf{R}$ at the time $t$ must be evaluated. (c) Interface between classical computing using the CP2K package and quantum emulators Qulacs and FQE. The interface is designed to facilitate a two-way exchange between CP2K and quantum emulators such as FQE or Qulacs. Specifically, it enables the transfer of second-quantized Hamiltonians generated by CP2K to the quantum emulator. Subsequently, the reduced density matrices computed by the quantum emulator are returned to CP2K for further processing.(d) The workflow for the MD simulation using the interface. The colors of the borders correspond to classical (blue) and quantum (red) components, as shown in panel (c).
}
\end{figure*}

\section{Methods}

This section outlines the interface developed in our research, whilst reviewing a methodology for force calculations with quantum computing.

Quantum chemistry calculations on quantum computers typically begin with a Hartree-Fock calculation performed on conventional quantum chemistry software. Then, a second-quantized fermionic Hamiltonian of the following form is computed:
\begin{align}
H = \sum_{p,q}h_{pq}c^{\dagger}_{p}c_{q} + \frac{1}{2}\sum_{p,q,r,s}h_{pqrs}c^{\dagger}_{p}c^{\dagger}_{r}c_{s}c_{q}
\end{align}
Here, $h_{pq}$ and $h_{pqrs}$ represent one-electron and two-electron integrals, respectively. The indices $p$, $q$, $r$, and $s$ denote molecular orbitals. $c^{\dagger}_{p}$ and $c^{\dagger}_{q}$ are fermionic creation operators, whilst $c_{r}$ and $c_{s}$ are annihilation operators.

Quantum computers operate on spin-1/2 systems and cannot directly manipulate fermionic operators. A fermionic Hamiltonian, therefore, must be transformed into a qubit-representation Hamiltonian $H^{\text{qubit}}$ using a fermion-to-spin mapping such as the Jordan-Wigner transformation~\cite{jordan1993paulische}. A qubit Hamiltonian $H^{\text{qubit}}$ is written as 
\begin{align}
H^{\text{qubit}} = \sum_{P} h_{P}P
\end{align}
where $P$ represents the tensor product of Pauli operators and $h_{P}$ are the corresponding coefficients.

Once $H^{\text{qubit}}$ is obtained, various quantum algorithms can be employed to prepare the wavefunction. For instance, the QPE can directly yield the eigenvalues and eigenvectors of $H^{\text{qubit}}$. Alternatively, the VQE can approximate these values variationally.

For variationally calculated wavefunctions, the Hellmann-Feynman theorem allows us to compute forces as the expectation value of the Hamiltonian differentiated with respect to nuclear coordinates. However, when employing active space approximations, the Hellmann-Feynman theorem no longer holds. In such cases, orbital response terms must be calculated by solving the Coupled Perturbed Hartree-Fock (CPHF) equations.~\cite{osamura1982generalization}

In either scenario, the calculation requires the one-particle reduced density matrix (1-RDM) $d^{\text{full}}_{pq}$, the two-particle reduced density matrix (2-RDM) $D^{\text{full}}_{pq,rs}$, and the nuclear coordinate derivatives of one- and two-electron integrals. The latter relies solely on classical computation, whilst RDMs are obtained by measuring the expectation values of operators $c^{\dagger}_{p} c_{q}$ and $c^{\dagger}_{p} c^{\dagger}_{r} c_{s} c_{q}$ for the quantum-computed state $|\Psi\rangle$. 

Our implementation, as illustrated in Figure 1(c), employs CP2K to generate the second-quantized Hamiltonian for the active orbital space. This Hamiltonian is then passed to either the Qulacs quantum circuit simulator~\cite{Suzuki_2021} or the Fermionic Quantum Emulator (FQE)~\cite{rubin2021fermionic}. These tools generate the 1-RDM and 2-RDM, which are subsequently returned to CP2K. Within CP2K, the second-quantized Hamiltonian is stored in the FCIDUMP format, which is then read by Qulacs or FQE. When using Qulacs, the OpenFermion package~\cite{mcclean2020openfermion} is utilized to convert the Hamiltonian (operator) from the fermionic form to a qubit form. Throughout this research, we exclusively employed the Jordan-Wigner transformation. For AIMD simulations, as shown in Figure 1(d), this force calculation is followed by an update of atomic positions and velocities. The process then loops back to the initial Hartree-Fock step, repeating the cycle as necessary.

The diagonalization of the second-quantized Hamiltonian was performed using both the VQE implemented in Qulacs and FQE. For the VQE ansatz, we adopted the Unitary Coupled Cluster with Singles and Doubles (UCCSD)~\cite{kutzelnigg1982quantum,hoffmann1988unitary,bartlett1989alternative,taube2006new,romero2018strategies, Anand_2022, Lim_2022}. For large-scale systems, an active space approximation was employed, restricting the electronic orbital degrees of freedom to regions critical for reactivity. The active space was selected by choosing an equal number of occupied and unoccupied orbitals near the Fermi level. When diagonalizing a second-quantized Hamiltonian with a 4-electron 4-orbital active space using VQE and FQE, the notations UCCSD(4$e$,4$o$) and FQE(4$e$,4$o$) are used, respectively.

\section{\label{sec:level1}Results and Discussion}

This section presents the results of the simulations conducted using the developed interface with the VQE and FQE methods under the active space approximation. We demonstrated the application of our approach to representative condensed matter systems, as follows. Section II.A examines water clusters, Section II.B investigates H$_2$O on the Si(001) surface, Section II.C explores liquid water, and Section II.D analyzes chorismate mutase. For the systems discussed in Sections II.A, II.B, and II.D, we performed geometry optimizations. Whereas, for the system in Section II.C, AIMD simulations were conducted to evaluate the oxygen-oxygen (O-O) radial distribution function (RDF), which is a key statistical property of liquid water. Further, in Section II.E, we assess the computational efficiency of the proposed interface. This was done by examining the system-size dependence of the VQE calculation time using the UCCSD ansatz for liquid water simulations at a density of \SI{1}{g/cm^3} for various sizes of the PBC box. Computational details can be found in Appendix~\ref{app:B}.

\subsection{\label{sec:level2}Water clusters}

\begin{figure}[]
\includegraphics{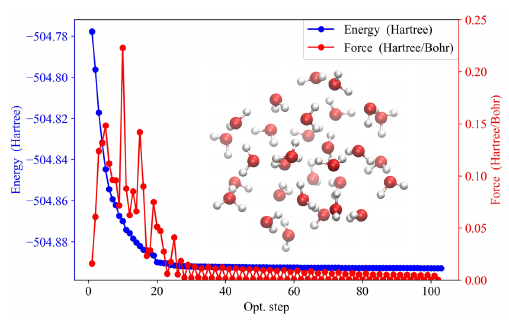}
\caption{\label{fig:Fig2} Geometry optimization of the water cluster $\mathrm{(H_{2}O})_{30}$ at the UCCSD(2$e$,2$o$) level. The left axis and blue dots represent the total energy. Whereas, the right axis and red dots illustrate the force as a function of the optimization step (opt. step). 
}
\end{figure}

Geometry optimizations were performed for the water cluster, that is, the water n-mer (H$_2$O)$_n$ (n=2 $\cdots$ 30). Figure 2 shows the geometry optimization at the UCCSD(2$e$,2$o$) level starting from (H$_2$O)$_{30}$, which was preoptimized using the HF method. The energy decreased uniformly, and after approximately the 20th optimization step, the force decreased with oscillations below 0.05 Hartree/Bohr. In the 102nd optimization step, the force reduced to below the convergence threshold of $0.45\times10^{-3}$ Hartree/Bohr, and convergence was achieved. The optimized structure of (H$_2$O)$_{30}$ is shown in Figure 2. The binding energies per $\mathrm{H_2O}$ of the water cluster (H$_2$O)$\mathrm{_n}$ (n=2 $\cdots30$) before and after geometry optimization are shown in Figure 3. The binding energies of all the water clusters increased with geometry optimization, thereby confirming the possibility of identifying stable structures via geometry optimization. As presented in Appendix B, the structures before and after optimization for n=2–30 are very similar. However, the binding energies exhibited a change greater than 1 kcal/mol, which is beyond the chemical accuracy. Thus, the equilibrium geometry must be determined for calculating binding energies or other relative chemical quantities with chemical accuracy. 

In the case of (H$_2$O)$\mathrm{_2}$, the binding energy before geometry optimization was negative (-0.50 kcal/mol/H$_2$O). This implied that the water molecule was more stable when it existed by itself. The binding energy in the optimized geometry was 2.48 kcal/mol/H$_2$O, which indicated stabilization via hydrogen bonding. The full configuration interaction calculation for the water dimer at the complete basis set limit yielded 5.1 kcal/mol (2.55 kcal/mol/H$_2$O)~\cite{feller1992application} and experimental measurements of 5.4 ± 0.7 kcal/mol (2.7 ± 0.35 kcal/mol/H$_2$O). These values are consistent with the present calculation. The energy difference can be attributed to the size of the chosen basis set 6-31G(d) not sufficiently large, the insufficient size of the active space and the effect of the zero-point energy. 

\begin{figure}[h]
\includegraphics{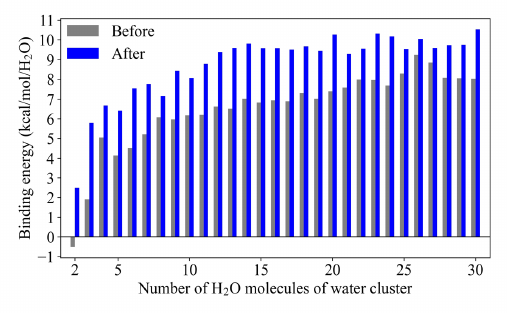}
\caption{\label{fig:Fig3}Binding energy per water molecule of water clusters $\mathrm{(H_{2}O})_\mathrm{n}$ with n=2–30 before (gray) and after (blue) geometry optimization at the UCCSD(2$e$,2$o$) level.}
\end{figure}

\subsection{\label{sec:level2}Defect on semiconductor surface}

Next, to apply the surface/interface systems, we calculated the adsorption energies for the dissociative adsorption of H$_2$O molecule on the Si(001) surface. In particular, we calculated the adsorption energy of the adsorption structure known as a type-C defect ~\cite{yu2008extrinsic}. The Si(001) surface is stabilized by an alternating arrangement of inclined dimers that reconstruct a $c$(4$\times$2) periodic structure ~\cite{ramstad1995theoretical, hamers1986scanning, shirasawa2006structural}. Scanning tunneling microscopy measurements indicated that residual water molecules were dissociatively adsorbed onto the Si(001)-$c$(4$\times$2) surface. The dominant structures were H-Si and HO-Si bonds formed on the same side Si atoms for neighboring Si dimers~\cite{yu2008extrinsic}. For modeling the Si(001) surface, single-point calculations were performed at the FQE(4$e$,4$o$) level by varying the diamond Si lattice constant (Figure 4). The most stable lattice constant 5.4 Å was consistent with the experimental value of 5.43 Å. Thus, this lattice constant was used to model the Si(001)-$c$(4$\times$2) surface in the hydrogen-terminated five-layer slab model, as shown in the bird’s eye view in Figure 5.

\begin{figure}[]
\includegraphics{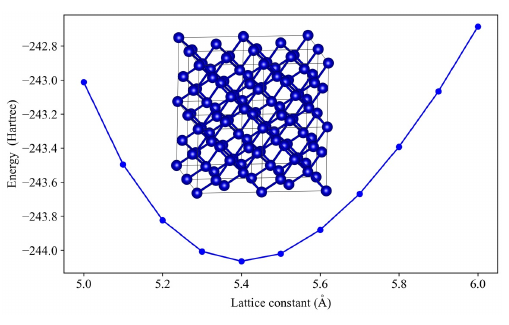}
\caption{\label{fig:Fig4}Energy versus lattice constant for bulk silicon. The calculated energy is shown as a function of the lattice constant. The minimum energy is observed at a lattice constant of 5.4 Å, which is the value adopted for subsequent modeling of the Si surface. The inset illustrates the crystal structure of bulk silicon used in the calculations.
}
\end{figure}

\begin{figure}[b]
\includegraphics{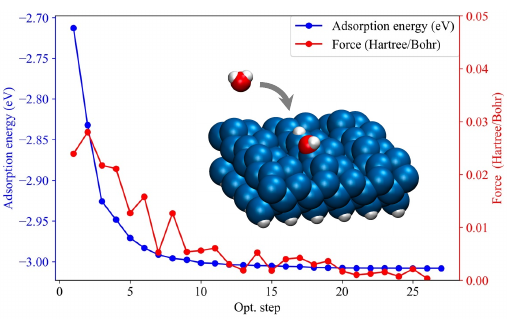}
\caption{\label{fig:Fig4}Geometry optimization of the dissociative adsorption of a $\mathrm{H_2O}$ molecule on the Si(001)-c(4$\times$2) reconstruction surface at the UCCSD(4$e$,4$o$) level.  The red axis and points on the left indicate the adsorption energies calculated using the energies of water molecules and Si(001)-$c$(4$\times$2) clean surfaces. The blue axis and points on the right present the force as a function of the optimization step.
}
\end{figure}

Figure 5 shows the geometry optimization of the dissociative adsorption of H$_2$O on the Si(001) surface at the UCCSD(4$e$,4$o$) level. Similar to the molecular system, the periodic surface system demonstrated an appropriate decrease in energy at each geometry optimization step. The force converged below a threshold value of $0.45\times10^{-3}$ Hartree/Bohr at the 26th step. Further, the adsorption energy after geometry optimization was -3.11 eV. The DFT calculation using the 6-layer slab model with plane-wave basis functions and PBE functionals estimated the adsorption energy to be -2.12 eV~\cite{yu2008extrinsic}. However, our calculations overestimated the adsorption energy by approximately -1.01 eV. To determine the reason for this difference, the stable structure was calculated using DFT calculations with the PBE functional. This yielded -2.72 eV, which was close to the value of -2.12 eV. The remaining energy difference cfould be attributed to the difference in the basis functions, GPW basis functions (6-31G(d), Gaussian basis sets with auxiliary plane-wave basis functions), and plane-wave basis functions. To the best of knowledge, no experimental data have been reported, and quantum chemical calculations of adsorption energies are scarce. Thus numerical experiments on a wide range of surface adsorption systems using various levels of theory are required to realize theoretical prediction accuracy compared to the experimental measurements.

\subsection{\label{sec:level2} Liquid water}

\begin{figure}
\includegraphics{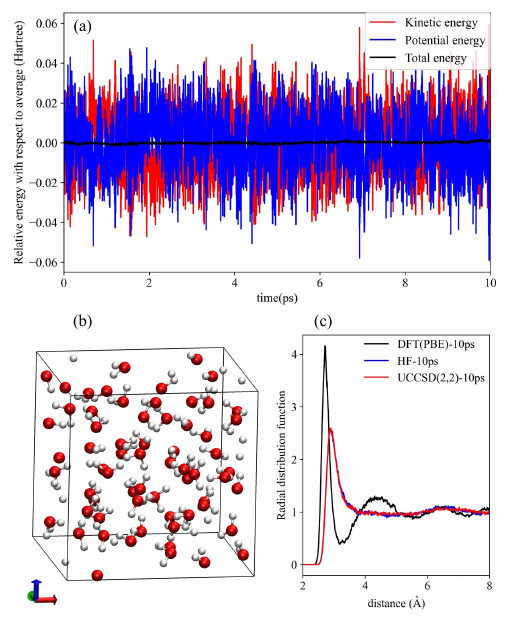}
\caption{\label{fig:Fig5}Results of a 10 ps NVE simulation of liquid water at 300 K at the UCCSD(2$e$,2$o$) level. (a) Plot of kinetic energy (red line) and potential energy (blue line) and their sum (sum) with respect to their respective averages (b) Snap-shot of the liquid water at 10 ps with PBC box. (c) Radial distribution function (RDF) of oxygen atoms calculated via NVE simulations of 10 ps for a total of 20,000 trajectories using DFT (black), HF (blue), and UCCSD(2$e$,2$o$) (red), respectively.}
\end{figure}

The results of a 10 ps UCCSD(2$e$,2$o$) NVE AIMD simulation of liquid water at 300 K are presented in Figure 6. Figure 6(a) shows the obtained results, and Figure 6(b) shows the trajectories at 10 ps, including the periodic boundary box. The conservation of total energy, as indicated by the black line, confirmed the accurate evaluation of the forces through the analytical gradient calculations. Subsequently, we examined the RDF of the oxygen atoms in the 10 ps NVE simulations performed using the DFT, HF, and UCCSD(2$e$,2$o$) methods (Figure 6(c)). The RDFs obtained from the HF and UCCSD(2$e$,2$o$) simulations exhibited strikingly similar profiles with two major peaks. This similarity may be attributed to the employment of a relatively small active space. This yielded accuracy levels comparable to the HF calculations because the active space is tiny relative to the full system, which contains 512 electrons distributed over 1,152 orbitals. These results were consistent with those of a previous study that used the HF potential~\cite{swaminathan1977theoretical}. In contrast, the RDF derived from the DFT simulations exhibited prominent peaks that spanned from the first to the third hydration sphere, thereby successfully reproducing the number of major peaks observed in the experimental measurements~\cite{skinner2013benchmark}. These findings demonstrate that HF and UCCSD with limited active spaces do not accurately replicate the RDF of oxygen atoms in liquid water.


\begin{figure}
\includegraphics{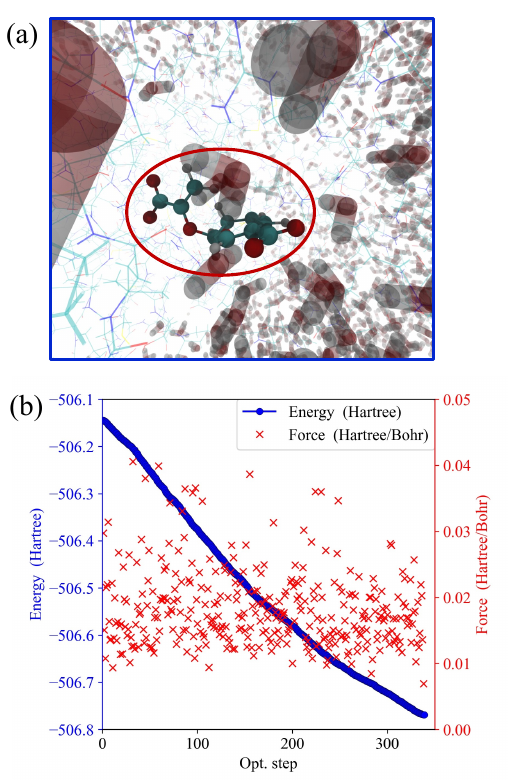}
\caption{\label{fig:Fig6}Chorismate mutase. The chorismate circled in red is the QM region and the others are the MM region. (b) Geometry optimization of chorismate mutase. The blue axis and blue dots on the left indicate the energy, and the red axis and red × symbol indicate the force on the right of the QM region.}
\end{figure}

\subsection{\label{sec:level2} Chorismate mutase enzyme}

Chorismate mutase is an essential enzyme found in various organisms, including plants and bacteria~\cite{kast1996chorismate}. It plays a crucial role in specific physiological functions. It functions as a catalyst in certain biochemical reactions and facilitates the conversion of shikimic acid (chorismate) to prephenate, which is a significant step in the aromatic amino acid biosynthesis pathway. In this study, we took the first step towards simulating this enzyme reaction by performing a geometry optimization of the reactant, chorismate, within the framework of quantum mechanics/molecular mechanics (QM/MM)~\cite{senn2009qm}. The QM region comprised the substrate chorismate, whereas the MM region represents the enzyme environment (see Figure 7(a)). Detailed information on the model setup and the classical force field settings can be found in~\cite{cp2k_chorismate}.

This study employed the UCCSD(4$e$,4$o$) level of theory for QM calculations. The progress of geometry optimization is illustrated in Figure 7(b). The geometry optimization using QM/MM proceeded stably. The energy decreased consistently throughout the optimization process. Although the change in forces did not stabilize until immediately before converging to 0.01 Hartree/Bohr, this behavior was attributable to the perturbation of the QM region owing to interactions with the MM region. However, with the progression of the optimization, the fluctuation diminished. We anticipate that implementing more stringent convergence criteria for the forces will result in smaller fluctuations. Thus, these results demonstrate the feasibility of the VQE simulation within the QM/MM framework.

\subsection{\label{sec:level2} Computational efficiency}

\begin{figure}
\includegraphics{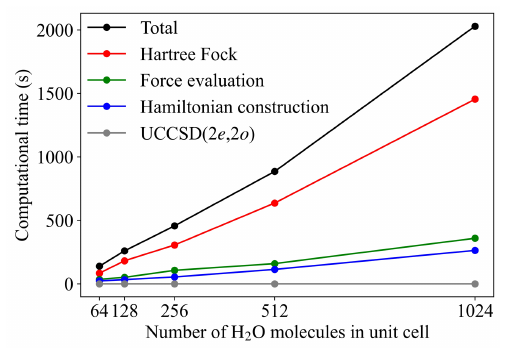}
\caption{\label{fig:Fig7}Single-point calculation benchmarks for bulk water of \SI{1}{g/cm^3} at UCCSD(2$e$,2$o$) level, modeled considering different simulation box sizes of 64, 128, 256, 512, and 1024 $\mathrm{H_2O}$ molecules. Different stages of computations are distinguished: total computation time (black circles), HF computation time (red circles), analytic derivative computation time (green circles), second-quantized Hamiltonian construction time (blue circles) and UCCSD(2$e$,2$o$) calculations. All calculations are performed using 4 Intel(R) Xeon(R) Platinum 9242 CPUs with 96 CPU cores in Message Passing Interface (MPI) parallel, for a total of 384 CPU cores.}
\end{figure}

We evaluated the performance of the VQE simulation interface developed in this study. Specifically, we focused on the system size dependence of the UCCSD(2$e$,2$o$) level calculations for liquid water at 300K. Figure 8 illustrates the scalability of various components involved in the UCCSD(2$e$,2$o$) calculations, including the HF calculations, second-quantized Hamiltonian construction, and the evaluation of forces using analytical differentiation as a function of the number of water molecules. HF calculations performed using CP2K exhibited approximately linear scaling with an increase in the number of H$_2$O molecules in the unit cell. This was consistent with the linear scaling reported in the benchmarks of Kühne et al. for AIMD calculations of liquid water~\cite{guidon2008ab}. Similarly, the costs associated with the second-quantized Hamiltonian construction and force evaluation scaled linearly with the number of molecules. The computation time for the UCCSD calculations was significantly shorter than that for the other processes performed on the CP2K side. This is because the active space for execution on a quantum computer is limited to four qubits. In the future, as the scale of accessible quantum devices increases, it is expected that the degrees of freedom of orbitals handled in quantum calculations will increase, which would facilitate the addressal of more meaningful tasks.

\section{\label{sec:level1}Conclusion}

This study developed an interface between CP2K, a linear-scaling quantum chemical calculation package, and Qulacs and FQE, which are state vector-based quantum simulators, to facilitate efficient quantum-classical hybrid calculations for realistic chemical problems. The developed interface facilitated the evaluation of analytical gradients with respect to the atomic coordinates using the 1-RDMs and 2-RDMs obtained from the VQE simulations. Water clusters, liquid water, dissociative adsorption structures of water molecules on the reconstructed Si(001)-$c$(4$\times$2) surface, a well-known defect in silicon wafers, and chorismate mutase, an important enzymatic reaction in biological systems, were selected to conduct simulations on realistic materials. The computational efficiency of the developed interface was evaluated at the UCCSD(2$e$,2$o$) level for liquid water with a density of \SI{1}{g/cm^3} and a periodic boundary condition box containing 64–1024 molecules. The developed interface appropriately elicited a near-linear performance from CP2K, which facilitated the second-quantized Hamiltonian construction and analytical force evaluation at a near-linear cost with respect to the number of molecules and simulation box size. 

However, issues regarding accuracy were observed. The RDF of water at 300K obtained at the UCCSD(2$e$,2$o$) level was almost identical to that at the HF level. The adsorption energy of H$_2$O on the Si(001)-$c$(4$\times$2) surface differed significantly from that obtained using DFT. These accuracy issues suggest that the orbital degrees of freedom for electron correlation included in quantum calculations were insufficient for large systems involving tens to thousands of atoms. These problems may be resolved by employing several sophisticated quantum-classical hybrid algorithms. For example, the methods proposed by Erhart et al.~\cite{erhart2024coupled} and Scheurer et al.~\cite{scheurer2023tailored} efficiently solved the active space problem in quantum algorithms, and incorporated dynamic correlations using methods such as CCSD. Furthermore, in molecular adsorption systems, improvements can be achieved through the application of quantum embedding methods~\cite{lee2019projection, battaglia2024general} that handle only the region close to the adsorption site with quantum calculations while treating the remaining using DFT. Using the interface we developed to implement these quantum-classical hybrid algorithms, it is expected that the computational burden on classical computing can be minimized, thereby enabling efficient quantum-classical hybrid algorithms.

\begin{acknowledgments}
This project was supported by funding from the MEXT Quantum Leap Flagship Program (MEXTQLEAP) through Grant No. JPMXS0120319794, the JST COI-NEXT Program Grant No. JPMJPF2014, and JST ASPIRE Program JPMJAP2319. The completion of this research was partially facilitated by the JSPS Grants-in-Aid for Scientific Research (KAKENHI) Grant Nos. JP23H03819.
We thank the Supercomputer Center, the Institute for Solid State Physics, the University of Tokyo for the use of the facilities. 
This work was also achieved through the use of SQUID at the Cybermedia Center, the University of Osaka.
\end{acknowledgments}

\appendix

\section{\label{app:A}Theoretical implementation}

The following has been written considering the spin-restricted case, although the code also applies to the spin-unrestricted case: All codes were implemented in a modified CP2K version 8.2 ~\cite{kuhne2020cp2k}

\subsection{\label{sec:level2}ENERGY EVALUATION}

The subspace Hamiltonian within the active space is computed using the closed-shell Fock operator:
\begin{eqnarray}
f_{xy}^c=h_{xy}+G({\bf{d}}^c)_{xy}, \\
V_{xy,zw}= (xy\vert zw)  
\end{eqnarray}
where $x, y, z$, and $w$ run over the active orbitals. $h_{xy}$ represents the one-electron Hamiltonian matrix elements in the basis of active orbitals. $G(\mathbf{d}^c)_{xy}$ denotes the Coulomb and exchange contributions from the density matrix of the closed (inactive) orbitals, $\mathbf{d}^c$. $f_{xy}^c$ is the Fock matrix within the active space, accounting for the interactions with the closed-shell electrons. $V_{xy,zw} = (xy|zw)$ are the two-electron repulsion integrals between active orbitals. The two-electron integrals $V_{xy,zw}$ are computed by expanding each product of active orbital pairs $\phi_x \phi_y$ into plane waves (PWs), which is efficient for periodic systems. 

An external active-space solver is used to obtain the correlation energy and the correlation contribution to 1-RDM and 2-RDM within the active space. We have implemented interfaces for both the FQE~\cite{rubin2021fermionic} and Qulacs~\cite{Suzuki_2021}, which compute
\begin{eqnarray}
&&E^c=E^{\mathrm{full}}-E^{\mathrm{HF}}, \\
&&d_{xy} = d_{xy}^{\mathrm{full}}-d_{xy}^{\mathrm{HF}},  \\
&&D_{xy,zw}=D_{xy,zw}^{\mathrm{full}}-D_{xy,zw}^{\mathrm{HF}},
\end{eqnarray}
where $E^{\mathrm{full}}$ is the total energy obtained from the full configuration interaction (FCI) calculation within the active space. $E^{\mathrm{HF}}$ is the Hartree-Fock (HF) energy recomputed within the active space using the active-space solver. $d_{xy}^{\mathrm{full}}$ and $D_{xy,zw}^{\mathrm{full}}$ are the 1-RDM and 2-RDM from the FCI calculation. $d_{xy}^{\mathrm{HF}}$ and $D_{xy,zw}^{\mathrm{HF}}$ are the corresponding HF 1-RDM and 2-RDM within the active space. $E^\mathrm{c}$ is the correlation energy within the active space. $d_{xy}$ and $D_{xy,zw}$ are the correlation contributions to the 1-RDM and 2-RDM, respectively. The integrals are passed to this solver using so-called FCIDUMP format and the resulting energy, 1-RDM, and 2-RDM are passed back in the same format. The total energy is expressed as follows:
\begin{equation}
E=E^{\mathrm{ref}}+\frac{1}{2}\sum_{xy} f_{xy}^cd_{xy}+\frac{1}{2}\sum_{xy,zw}V_{xy,zw}D_{xy,zw} 
\label{eq:6}
\end{equation}
$E^\mathrm{ref}$ is the reference HF energy, which should not be confused with $E^\mathrm{HF}$. In periodic cases, the plane-wave expansion involves reciprocal lattice vectors $\mathbf{g}$. The correlation contributions are computed under the approximation that the contributions from the $\mathbf{g} = \mathbf{0}$ reciprocal lattice vector are neglected. This is a common approximation in plane-wave-based calculations to avoid singularities associated with the long-range Coulomb interaction at $\mathbf{g} = \mathbf{0}$.

\subsection{\label{sec:level2}LAGRANGIAN AND MULTIPLIERS}
To discuss nuclear gradient formulation, it is worth noting that the total Lagrangian is expressed as

\begin{eqnarray}
L&&=E+\sum_{ai}Z_{ai}\frac{\partial E_{\mathrm{ref}}}{\partial \kappa_{ai}}+\frac{1}{2}\sum_{rs}X_{rs}({\bf{C^{\dagger}SC-1}})_{rs} \nonumber\\
&&+\sum_{i}\sum_{x^o}z_{ix^o}f_{ix^o}+\sum_a\sum_{x^v}z_{ax^v}f_{ax^v}.
\label{eq:7}
\end{eqnarray}
where $r$ and $s$ are the general molecular orbital (MO) indices and $\mathbf{C}$ is the MO coefficient matrix, $\mathbf{S}$ is the atomic orbital (AO) overlap integral. Hereafter, $i, j, k$, and $l$ denote the closed (inactive) orbitals, i.e., occupied orbitals that are outside of the active space. $a, b, c$, and $d$ denote virtual orbitals outside of the active space. This is performed on an AO basis. $x^{o}$ and $x^{v}$ denote the occupied and virtual 
parts of the active orbitals, respectively. In addition, $\kappa_{a i}$ represents the orbital rotation parameters between occupied orbitals $i$ and virtual orbitals $a$. The derivative ${\partial E_{\mathrm{ref}}}/{\partial \kappa_{a i}}$ denotes the SCF condition.

The Z-vector equation is derived by taking the derivative of the Lagrangian with respect to the orbital rotation between virtual and occupied orbitals and make it stationary:
\begin{equation}
Y_{bj}+({\bf{fz}+fz^{\dagger}})_{bj}+\sum_{ai}Z_{ai}\frac{\partial^2E_{\mathrm{scf}}}{\partial \kappa_{ai}\kappa_{bj}}=0 
\end{equation}
where $Y_{ai}={\partial E}/{\kappa_{ai}}$. The term with X does not appear here because it is symmetric (as opposed to anti-symmetric, such as $Z$). Because the SCF solution is a minimum, the second derivatives in the third term are positive definite; therefore, this equation is well-conditioned.

Once the RDMs are read from the file, they are used to compute $\mathbf{Y}$. In CP2K, we do not store virtual MOs explicitly. Therefore, the input to the Z vector equation is in the format of occupied MO coefficients where the first index is an AO index:
\begin{flalign}
Y_{\alpha i}&=\sum_{\beta}2P^v_{\alpha\beta}f^t_{\beta i}& \nonumber\\
&-\sum_{x^v}C_{\alpha{x^v}}\biggl[\sum_yf^c_{yi}d_{x^vy}+\sum_{y,zw}V_{yi,zw}D_{x^vy,zw}\biggl]& \\
Y_{\alpha x^o}&=\sum_\beta P^e_{\alpha\beta}\biggl[\sum_yf^c_{\beta y}d_{x^oy}+\sum_{y,zw}V_{\beta y,zw}D_{x^oy,zw}\biggl]&
\end{flalign}
with
\begin{eqnarray}
&&P^v={\bf{S}}^{-1}-\frac{1}{2}{\bf{d_{SCF}}} \\
\label{eq:11}
&&P^e={\bf{S}}^{-1}-\bf{C_{occ}C^{\dagger}_{occ}}
\label{eq:12}
\end{eqnarray}
where $\bf{C_{occ}}$ is the extended MO coefficients (including both inactive and active) and $\alpha$ and $\beta$ are the AOs. We also introduce the so-called total Fock operator,
\begin{equation}
f^t_{rs}=h_{rs}+G({\bf{d}^{\mathrm{ref}}+d})_{rs}.
\end{equation}
In CP2K the linear equation is multiplied from the left by the overlap matrix $\bf{S}^{-1}$. Therefore, it is not necessary to compute the inverse of the overlap matrix. Because we do not want to construct the two electron integrals with a virtual index, we use the following algorithm for the last term. First, we back-transform the two electron density to AOs for each $x^v$ and $y$, that is
\begin{equation}
D_{xy,zw}\equiv({\bf{d}}^{xy})_{zw} \rightarrow ({\bf{d}}^{xy})_{\alpha\beta}.
\end{equation}
This density matrix is used to obtain the Coulomb operator using the code in the SCF, which is subsequently transformed into MOs. For each $x$ and $y$, we compute
\begin{equation}
\sum_{zw}V_{yi,zw}D_{x^v,zw}=\sum_gJ({\bf{d}}^{x^vy})_g\phi_y(g)\phi_i(g)
\end{equation}
where $J_g$ is the Hartree potential at grid $g$. The following quantities are introduced in the code:
\begin{equation}
g_{\alpha x}=\sum_yf^c_{\alpha y}d_{yx}+\sum_y\sum_gJ({\bf{d}}^{xy})_g\phi_\alpha(g)\phi_y(g)
\end{equation}
Using this quantity, Eqs. (A9) and (A10) can be simplified to
\begin{eqnarray}
Y_{\alpha i}&&=\sum_\beta2P^v_{\alpha\beta}f^t_{\beta i}-\sum_{x^v}C_{\alpha x^v}g_{ix^v} \\
\label{eq:17}
Y_{\alpha x^o}&&=P^e_{\alpha \beta g \beta x^o}
\label{eq:18}
\end{eqnarray}
The multipliers $z$ associated with the active orbital selection using canonical orbitals are computed as follows:
\begin{equation}
z_{ij}=-\frac{Y_{ij}-Y_{ji}}{f_{ii}-f_{jj}}
\label{eq:19}
\end{equation}
Because CP2K implementation does not explicitly compute the virtual orbitals, this is solved iteratively for the virtual part of $\mathbf{z}$. After obtaining $\mathbf{z}$, Eq. (8) is solved iteratively to obtain $\mathbf{Z}$.

\subsection{\label{sec:level2}RELAXED DIPOLE MOMENT}
Once the $Z$ vector is obtained, the total relaxed density matrix can be computed as
\begin{equation}
\tilde d_{rs} = d_{rs}^{\mathrm{ref}}+d_{rs}+d_{rs}^z
\label{eq:20}
\end{equation}
where we introduced ${\bf{d}}^z={\bf{Z}}+\bf{z}$, with $\mathbf{Z}$ and $\mathbf{z}$ being the Lagrange multipliers associated with the constraints in the Lagrangian. The relaxed density matrix is defined as the derivative of the Lagrangian
\begin{equation}
\tilde d_{rs} = \frac{\partial L}{\partial h_{rs}}.
\label{eq:21}
\end{equation}
The dipole moment is calculated using this quantity. Under an external electric field, E, the Hamiltonian becomes $H \rightarrow H + \sum_{rs}\mu_{rs}{\bf{E}}_{rs}$, (i.e., the dipole operator appears whenever $h$ appears), therefore,
\begin{equation}
\mu=\frac{\partial L}{\partial \bf{E}}\bigg\vert _{{\bf{E}}=0}=\sum_{\alpha\beta}M_{\alpha\beta}\tilde d_{\alpha\beta}    
\label{eq:22}
\end{equation}
where we implicitly back-transform the density matrix into AO, and $M_{\alpha \beta}$ are the dipole moment integrals between atomic orbitals. This dipole moment agrees exactly with the finite difference obtained using the finite $\mathbf{E}$.

\subsection{\label{seq:level2}EVALUATION OF NUCLEAR GRADIENTS}
Once the relaxed density matrices are computed, the nuclear gradients are evaluated as follows.

\subsubsection{\label{sec:level3}Relaxed 1-RDM}
Using the total density matrix defined in Eq. (A20), the contribution to the nuclear gradients from the one-body part of the Hamiltonian can be expressed as
\begin{equation}
\frac{dE}{dR_A}=\frac{\partial L}{\partial R_A} \leftarrow 2h_{rs}^{R_A}\tilde d_{rs}
\label{eq:23}
\end{equation}
where $h_{r s}^{R_A} = \frac{\partial h_{r s}}{\partial R_A}$ represents the derivative of the one-electron Hamiltonian matrix elements with respect to the displacement of atom $A$ in the Cartesian direction $R$.
In practice, however, because nuclear attraction is handled together with the Hartree potential in CP2K, special care must be taken (see below).
\subsubsection{\label{sec:level3}Non-separable part of 2-RDM}
The 2-RDM from the active-space solver is contracted to the two-electron gradient integrals and contributes to the nuclear gradient as follows:
\begin{equation}
\frac{\partial L}{\partial R_A} \leftarrow 2\left(i_{A_R} j \mid k l\right) \Gamma_{i j, k l}
\label{eq:24}
\end{equation}
where $R$ is the Cartesian direction, and $A$ labels the atoms. $i_{A_R}$ is the derivative of MO $i$ with respect to the nuclear displacement (computed using the derivative basis functions and the same MO coefficients). $\Gamma_{i j, k l}$ is the 2-RDM element in the MO basis. This contribution is evaluated using the Hartree potential $J(\mathbf{d}^{x y})_g$ used to compute the intermediate quantities in Eq.~{(A17)} for each $x$ and $y$ pair: 
\begin{eqnarray}
    \frac{\partial L}{\partial R_A} && \leftarrow \sum_{z w}\left(x_{A_R} y \mid z w\right) D_{x y, z w} \nonumber\\
    && =\sum_g J\left(\mathbf{d}^{x y}\right)_g \frac{\partial \phi_x(g)}{\partial R_A} \phi_y(g) \nonumber\\
    && =\sum_g \sum_{\alpha \beta} \sum_g J\left(\mathbf{d}^{x y}\right)_g \frac{\partial \phi_\alpha(g)}{\partial R_A} \phi_\beta(g) T_{\alpha \beta}^{x y} 
\label{eq:25}
\end{eqnarray}
where $\phi_x(g)$ is the value of the active molecular orbital $x$ at grid point $g$, and $\phi_\alpha(g)$ is the atomic orbital (AO) basis function at grid point $g$. $T_{\alpha \beta}^{x y} = C_{\alpha x} C_{\beta y}$, where $C_{\alpha x}$ and $C_{\beta y}$ are the MO coefficients relating AOs to MOs, i.e., $\phi_x = \sum_\alpha C_{\alpha x} \phi_\alpha$. Note that the last equation can be evaluated by the standard SCF nuclear gradient code.

\subsubsection{\label{sec:level3}Separable part of 2-RDM}
The separable part of the 2-RDM is given by
\begin{eqnarray}
    \Gamma_{r s, t u}^{\mathrm{sep}} && =d_{r s}^{\mathrm{ref}}\left(2 \mathbf{d}^z+\mathbf{d}^{\mathrm{ref}}\right)_{t u} \nonumber\\
    && -\frac{1}{2} d_{r u}^{\mathrm{ref}}\left(2 \mathbf{d}^z+\mathbf{d}^{\mathrm{ref}}\right)_{t s} \nonumber\\
    && +d_{r s}^c d_{t u}-\frac{1}{2} d_{r u}^c d_{t s}
\label{eq:26}
\end{eqnarray}
where $d_{r s}^{\mathrm{ref}}$ is the 1-RDM from the reference Hartree-Fock calculation, $\mathbf{d}^z = \mathbf{Z} + \mathbf{z}$ (with $\mathbf{Z}$ and $\mathbf{z}$ being the Lagrange multipliers from the Z-vector equations). The indices $r, s, t, u$ run over molecular orbitals.
The density matrices are back-transformed into AOs before entering the gradient integral contraction. Using a short hand notation for the gradient contribution, we obtain
\begin{eqnarray}
f_{A_R}\left(\mathbf{d}^a, \mathbf{d}^b\right) && =\sum_{x y z w}\left(x_{A_R} y \mid z w\right) \nonumber\\
\quad \times \left(d_{x y}^a d_{z w}^b\right. && \left.+d_{z w}^a d_{x y}^b-\frac{1}{2} d_{x w}^a d_{z y}^b-\frac{1}{2} d_{z y}^a d_{x y}^b\right).
\label{eq:27}
\end{eqnarray}
where $x, y, z, w$ are AO indices, $x_{A_R}$ denotes the derivative of AO $x$ with respect to the displacement of atom $A$ in Cartesian direction $R$, and $\left( x_{A_R} y \mid z w \right)$ are the derivative two-electron integrals involving the derivative of AO $x$.
The above expressions are further expressed as follows:
\begin{eqnarray}
\frac{\partial L}{\partial R_A} \leftarrow f_{A R}\left(\mathbf{d}^{\mathrm{ref}}, 2 \mathbf{d}^z+\mathbf{d}^{\mathrm{ref}}\right)+f_{A R}\left(\mathbf{d}^c, \mathbf{d}\right).
\label{eq:28}
\end{eqnarray}
In CP2K, the code for evaluating the separable contribution is complicated because the Hartree potentials are handled in a complex manner (particularly in the GAPW code). Therefore, we implemented these terms by taking advantage of the following relationship: dropping the subscript of $f$ for simplicity we obtain
\begin{flalign}
    & f\left(\mathbf{d}^a, \mathbf{d}^b\right)+f\left(\mathbf{d}^b, \mathbf{d}^a\right)& \nonumber\\
    & =f\left(\mathbf{d}^a+\mathbf{d}^b, \mathbf{d}^a+\mathbf{d}^b\right)-f\left(\mathbf{d}^a, \mathbf{d}^a\right)-f\left(\mathbf{d}^b, \mathbf{d}^b\right).&
\label{eq:29}
\end{flalign}
In the actual code, this idea is generalized such that the nuclear attraction potential which is handled together with the two-electron part, is appropriately handled. The working equation for the mixed term between ${\bf{d}}^{\mathrm{ref}}$  and ${\bf{d}}^z$ is expressed as
\begin{flalign}
    & f\left(-\varphi^{-1} \mathbf{d}^{\mathrm{ref}}+\varphi \mathbf{d}^z,-\varphi^{-1} \mathbf{d}^{\mathrm{ref}}+\varphi \mathbf{d}^z\right)-(1+\varphi) f\left(\mathbf{d}^z, \mathbf{d}^z\right)& \nonumber\\
    & -f\left(-\varphi^{-1} \mathbf{d}^{\mathrm{ref}},-\varphi^{-1} \mathbf{d}^{\mathrm{ref}}\right)+(1+\varphi) f(\mathbf{0}, \mathbf{0})&
\label{eq:30}
\end{flalign}
where $\varphi$ is the so-called golden ratio, $(1+\sqrt{5})/2$.

\subsubsection{\label{sec:level3}Overlap derivatives}
The contribution, which depends on the derivative of the overlap integrals is expressed as
\begin{eqnarray}
    \frac{\partial L}{\partial R_A} \leftarrow X_{rs}S_{r_{R_A}s} 
\label{eq:31}
\end{eqnarray}
where $X_{r s}$ is calculated from the symmetric part of the $\mathbf{Y}$ matrix, and $S_{r_{A_R}s}$ denotes the derivative of the overlap matrix elements with respect to the displacement of atom $A$ in Cartesian direction $R$.

\section{\label{app:B}Computational Details}
The calculations were performed using a customized CP2K version 8.2 in conjunction with the quantum computer emulator Qulacs. 
The computational conditions employed for each of the systems analyzed are as follows:

\begin{figure*}[]
\includegraphics{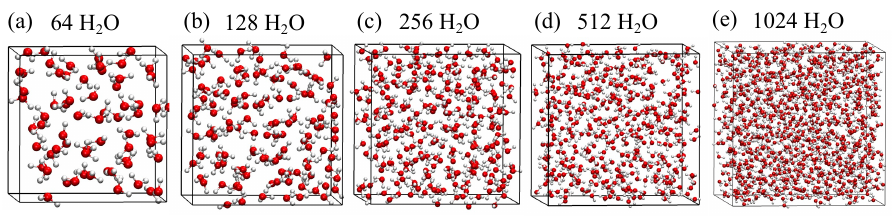}
\caption{\label{fig:FigS4}Liquid water unit cells with a density of \SI{1}{g/cm^3} used for benchmarking, as described in Section II.E. The unit cells contain (a) 64 H$_2$O, (b) 128 H$_2$O, (c) 256 H$_2$O, (d) 512 H$_2$O, and (e) 1024 H$_2$O molecules. Each snapshot represents the final configuration obtained after 1 ns of NVT equilibration at 300 K using the classical force field q-SPC/Fw.
}
\end{figure*}

\begin{figure*}[]
\includegraphics{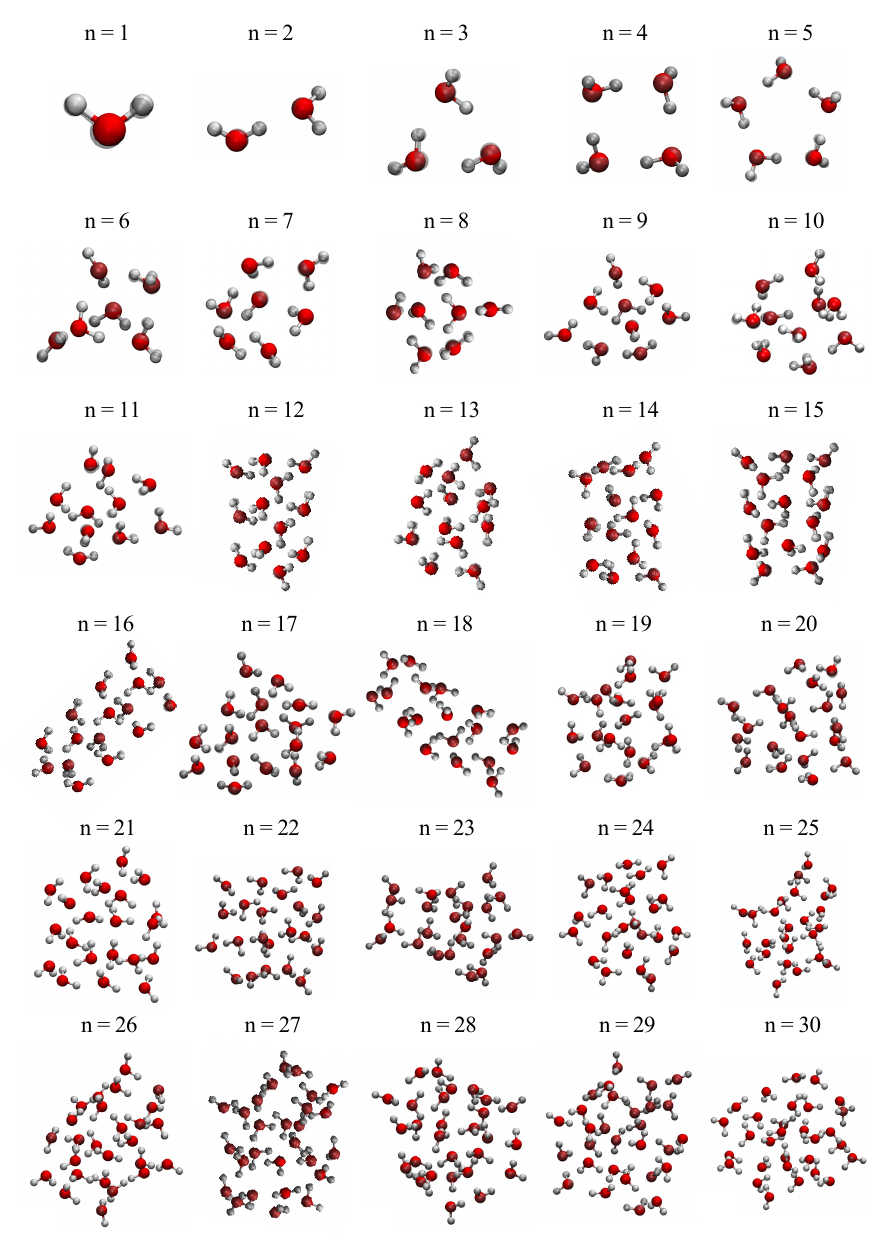}
\caption{\label{fig:FigS1} Optimized geometries of water clusters (n-mers) for n=1–30. Each subfigure displays the optimized geometry (with clear bonds and atoms) and the initial geometry used for the optimization (shown as shadows) using the UCCSD(2$e$,2$o$) calculations. 
}
\end{figure*}

\begin{figure*}[]
\includegraphics{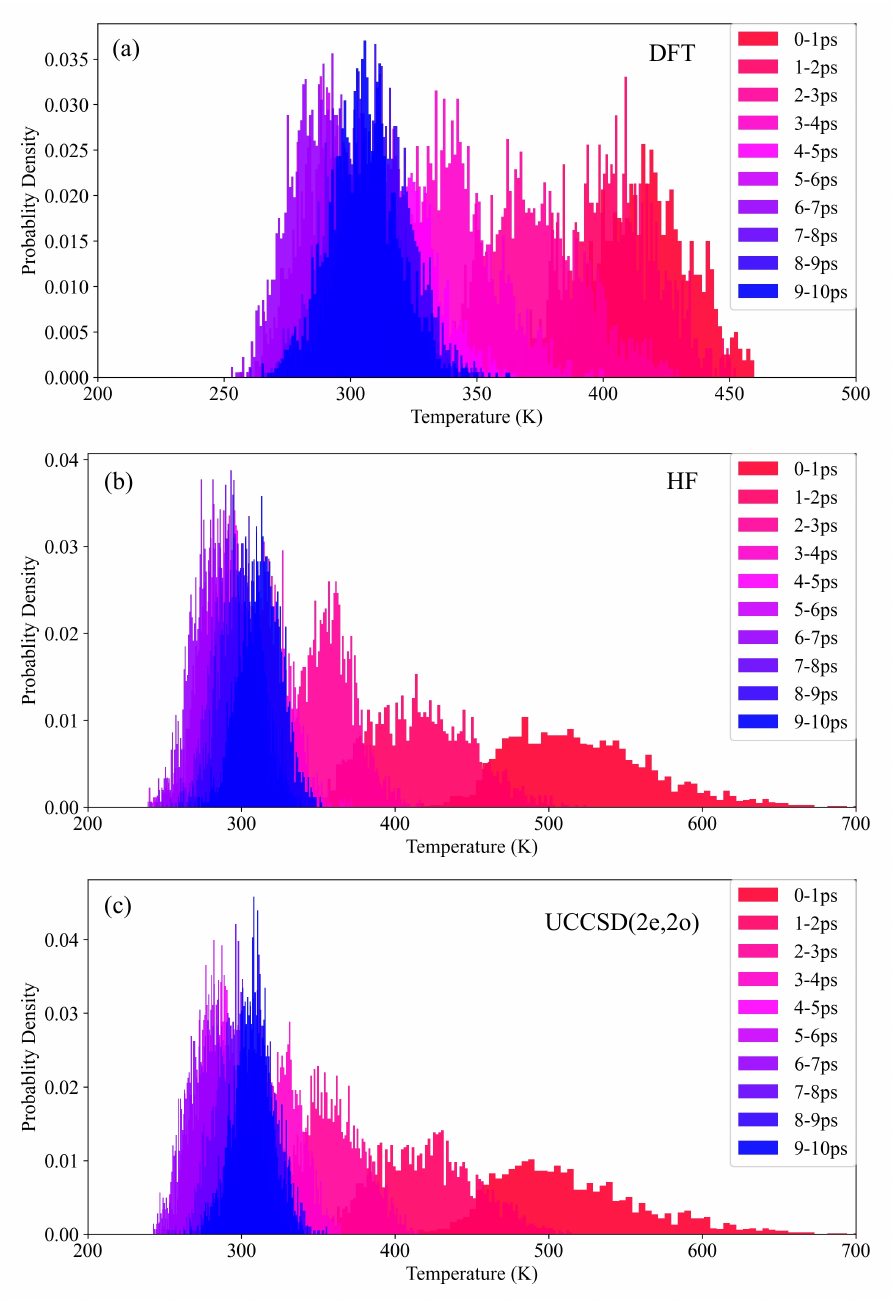}
\caption{\label{fig:FigS3} Temperature distribution changes during 10 ps NVT equilibration. The distributions are presented for every 1 ps interval. Each subplot represents a different computational method: (a) DFT, (b) HF, and (c) UCCSD(2$e$,2$o$). The x-and y-axes represent the temperature and probability density, respectively. The color gradient indicates the progression of time intervals, with each color corresponding to a specific 1 ps interval, as indicated in the legend.
}
\end{figure*}

\begin{figure*}[]
\includegraphics{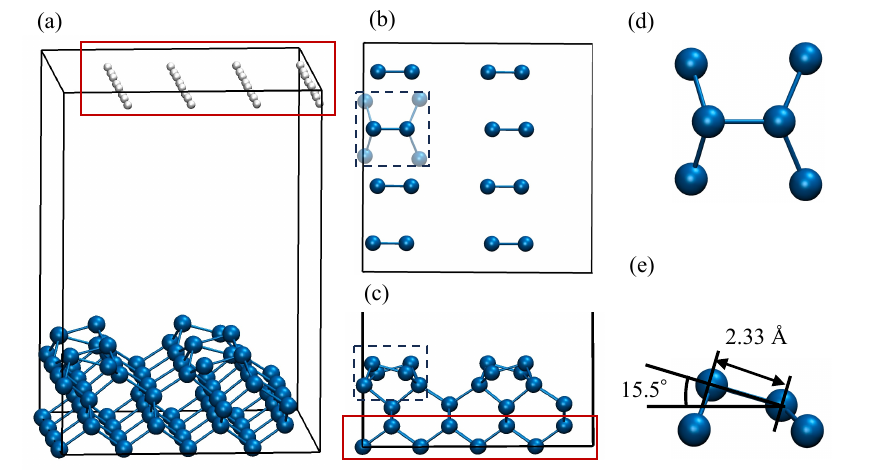}
\caption{\label{fig:FigS5} (a) Bird's eye view of the slab model of the Si(001)-c(4×2) reconstructed surface optimized at the UCCSD(4$e$,4$o$) level. The red box highlights the hydrogen termination for the fifth Si layer. (b) Top view and (c) side view of the Si(001)-c(4×2) reconstructed surface. In the top view, only the Si dimers of the reconstructed surface are shown for clarity. The black dashed box indicates the second layer of backbonding atoms (shown faintly) that bond with the Si dimers. In the side view, the red box indicates the fourth and fifth layers, where atomic positions are fixed during structural optimization, along with the hydrogen atoms in (a). (d) and (e) provide detailed views of the dashed box region in (c), with (e) presenting the Si dimer bond length and buckling angle.
}
\end{figure*}

\subsection{\label{sec:level2}Water cluster}

The initial geometries of the water clusters $\mathrm{(H_2O)_n}$ (n = 2–30),  comprising of n water molecules were obtained from n = 2–20~\cite{maheshwary2001structure} and n = 21–30~\cite{rakshit2019atlas}. The PBC box size was chosen to ensure a vacuum layer between clusters of less than 8 Å, specifically $20\times20\times20 \AA^3$. The HF calculations were performed using the GPW basis set under PBC. Further, the core electrons were treated with the corresponding Goedecker–Teter–Hutter (GTH) pseudopotential~\cite{goedecker1996separable}. The Gaussian-type basis function employed was 6-31G(d). The cutoff energy for the auxiliary plane-wave and Gaussian-type basis functions were set to 240 and 40 Ry, respectively. Four multigrid levels were utilized for efficient integral calculations. Wave function optimization employed the conjugate gradient method (CG), with an SCF threshold of $1.0\times10^{-8}$ Hartree. Further, orbital transformation (OT) was applied with an energy gap and step size of 0.08 Hartree. To enhance SCF convergence, the maximum number of loops for the outer SCF, which involved preconditioning at each step, was set to 10. In addition, the maximum number of loops for the inner SCF with a fixed preconditioner was set to 40. Moreover, to facilitate computational efficiency, two-electron integral (ERI) calculations were screened using the Schwarz inequality with a threshold below $1.0\times10^{-10}$.

Geometry optimization was performed using CG as the optimizer until the maximum force on an atom was less than $0.45\times10^{-3}$ Hartree/Bohr. The binding energy per water molecule in the water cluster $E_{\mathrm{bind}}$ is defined as
\begin{equation}  
E_{\mathrm{bind}} =\frac{\mathrm{n}E\mathrm{(H_2O)}-E\mathrm{(H_2O)}_\mathrm{n}}{\mathrm{n}}
\end{equation}
where $E\mathrm{(H_2O)}$ and $E\mathrm{(H_2O)_n}$ represent the total energies of a water molecule and water cluster, respectively.

\subsection{\label{sec:level2}Diamond Si crystal}
The diamond Si crystal was represented by a PBC box containing 64 atoms. HF calculations were performed employing the GPW basis set with core electrons treated using GTH pseudopotentials. The Gaussian-type basis function employed was 6-31G(d). The cutoff energy for the auxiliary plane-wave and Gaussian-type basis functions were set to 320 and 40 Ry, respectively. Four multigrid levels were utilized for efficient integral calculations. Wave function optimization employed the direct inversion in the iterative subspace (DIIS) method, with an SCF threshold of $1.0\times10^{-7}$ Hartree. Further, OT with a full single inverse was applied utilizing an energy gap and a step size of 0.08 Hartree. To enhance the SCF convergence, the maximum number of loops for the outer SCF, which involved preconditioning at each step, was set to 10. The maximum number of loops for the inner SCF with a fixed preconditioner was set to 40. For ERI calculations, the Schwarz inequality was utilized to screen values below $1.0\times10^{-8}$. In addition, a Truncated Coulomb operator was employed with a cutoff radius of 5 Å.

\subsection{\label{sec:level2}H$_2$O on Si(001) surface}
The Si(001)-c(4×2) reconstruction surface was modeled with a 4×4 size hydrogen-terminated 5-layer slab, where the vacuum layer was approximately 13 Å. The HF calculations were performed using the GPW basis set under PBC. The core electrons were treated with appropriate GTH pseudopotentials. The Gaussian-type basis function utilized was 6-31G(d). The cutoff energy for the auxiliary plane-wave and Gaussian-type basis functions were set to 400 and 60 Ry, respectively. To enhance the computational efficiency, four multigrid levels were employed for efficient integral calculations. Wave function optimization employed the DIIS method, with an SCF threshold of $1.0\times10^{-7}$ Hartree. Further, OT with a full single inverse was applied utilizing an energy gap and a step size of 0.08 Hartree. To improve the SCF convergence, the maximum number of loops for the outer SCF, which involved preconditioning at each step, was set to 10. The maximum number of loops for the inner SCF with a fixed preconditioner was set to 40. For ERI calculations, the Schwarz inequality was utilized to screen values below $1.0\times10^{-8}$. In addition,  a Truncated Coulomb operator was applied with a cutoff radius of 6 Å. Further, geometry optimization was performed using the conjugate gradient (CG) method as the optimizer until the maximum force on the adsorbed molecules and atoms up to the second layer on the surface was less than $0.45\times10^{-3}$ Hartree/Bohr. The adsorption energy of H$_2$O on the Si(001) surface $\mathrm{E_{ad}}$ was calculated using the following equation:
\begin{equation}  
E_{\mathrm{ad}} =E\mathrm{(H_2O/Si(001))}-E\mathrm{(H_2O)}-E\mathrm{(Si(001))}
\end{equation}
where $E\mathrm{(H_2O/Si(001))}$, $E\mathrm{(H_2O)}$, and $E\mathrm{(Si(001))}$ are the total energies of the H$_2$O/Si(001), H$_2$O, and Si(001), respectively.

\subsection{\label{sec:level2}Liquid water}

Liquid water with a density of \SI{1}{g/cm^3} was modeled in simulation boxes containing 64, 128, 256, 512, and 1024 $\mathrm{H_2O}$. NVT equilibration at 300 K was performed for each liquid water model using a Nose–Hoover thermostat at 10 ps. The NVT equilibration to 300 K at 1 ns was performed using AmberTools22~\cite{salomon2013overview} with the classical force field q-SPC/Fw~\cite{paesani2006accurate}. Final snapshots of each NVT equilibration are shown in Figure 9. NVT equilibration at \SI{300}{K} at \SI{10}{ps} was performed using DFT and HF. The step times were all set to 0.5 fs in this study. Subsequently, the 10 ps AIMD simulation in the NVE ensemble was performed based on the DFT and HF NVT equilibration results. Further, the UCCSD(2$e$,2$o$) NVE simulation was continued from the HF-level NVT equilibration.

The HF calculations were performed on a GPW basis under PBC, and the core electrons were treated with the corresponding GTH pseudopotential. The Gaussian-type basis function was 6-31G(d). The cutoff energy of the auxiliary plane-wave and Gaussian-type basis functions were set to 300 and 40 Ry, respectively. The number of multigrids was set to four for efficient integral calculations. Further, DIIS was adopted as the optimizer of the wave function, and the threshold of the SCF was set to $1.0\times10^{-7}$ Hartree. OT with a full single inverse preconditioner was applied with an energy gap and step size of 0.08 Hartree to improve the efficiency of the SCF calculation. To improve the convergence of the SCF, we set the maximum number of loops of the outer SCF, wherein the preconditioner was applied at each step, to 10. In addition, the maximum number of loops of the inner SCF, in which the preconditioner was fixed, was set as 40. Further, we screened ERI below $1.0\times10^{-10}$ using Schwarz's inequality to accelerate the computation. In the DFT calculations, the PBE functional~\cite{Perdew_1996} was selected as the exchange-correlation functional. The other calculation conditions were the same as those used for the HF calculation.

\section{\label{app:C}Geometries of water clusters}

Figure 10 shows the water clusters before and after the geometry optimization. No convergence to non-physical structures was observed. Thus, the geometry optimization based on the analytical gradients via CP2K, utilizing the quantum calculation results obtained through our developed interface, was confirmed to operate correctly.

\section{\label{app:D}NVT ensemble of liquid water}

Figure 11 shows the results of the NVT simulations used to determine the initial structure and velocities for the 10 ps NVE simulation of water at 300 K, as shown in Figure 6. The simulation procedures are detailed in Section V.D. In each case, the system equilibrated at approximately 300 K within 10 ps. The equilibration behavior at the UCCSD(2$e$,2$o$) level shown in Figure 10(c) resembled that of the HF method shown in Figure 10(b). This similarity was consistent with the RDFs for O-O obtained from the NVE ensemble, which were similar for both HF and UCCSD(2$e$,2$o$), as shown in Figure 6 (c).

\section{\label{app:E}Si(001)-c(4×2) reconstructed surface}

Figure 12 shows the unit cell of the Si(001)-c(4×2) surface slab model optimized at the UCCSD(4$e$,4$o$) level. Shirasawa et al. studied this surface using low-energy electron diffraction  experiments. They reported an Si dimer bond length of 2.4 ± 0.1 Å and a tilt angle of 18 ± 1°~\cite{shirasawa2006structural}. In our study, the optimized Si dimer bond length and tilt angle were 2.33 Å and 15.1°, respectively, as shown in Figure 12(e). These values were slightly smaller than the experimental values. This discrepancy may have resulted from the limited degrees of freedom of the orbitals used in the quantum calculations. Healy et al. applied quantum Monte Carlo methods to cluster models of a Si(001) surface and concluded that both static and dynamic correlations were essential for Si dimer buckling~\cite{PhysRevLett.87.016105}. We expect that simulations that can more efficiently and accurately incorporate electronic correlations, such as those using quantum embedding methods~\cite{lee2019projection} or large active spaces accessible with actual quantum devices, will achieve better consistency with experimental measurements.


\begin{verbatim}
\end{verbatim}

\nocite{*}

\bibliography{apssamp}

\end{document}